\documentclass[10pt,conference]{IEEEtran}
\usepackage{packages} 
\title{\toolname: A Platform for Engineering and Experimenting with Configurable Multi-Agent Architectures for Software Engineering}

\author{
\IEEEauthorblockN{Xiang Feng Ye Pan}
\IEEEauthorblockA{
Universitat Politècnica de Catalunya\\
Barcelona, Spain\\
xiangfeng.yepan@estudiantat.upc.edu
}
\and
\IEEEauthorblockN{Vincenzo De Martino}
\IEEEauthorblockA{
Universitat Politècnica de Catalunya\\
Barcelona, Spain\\
vincenzo.de.martino@upc.edu
}
\and
\IEEEauthorblockN{Matias Martinez}
\IEEEauthorblockA{
Universitat Politècnica de Catalunya\\
Barcelona, Spain\\
matias.martinez@upc.edu
}
}
\date{June 2026}

\begin{document}

\pagenumbering{arabic}
\pagestyle{plain}

\maketitle

\begin{abstract}
Experimenting with LLM-based multi-agent systems (MAS) involves an iterative cycle of modeling configurations, executing and evaluating them, comparing results, and refining system designs. Existing approaches provide limited end-to-end support for this experimentation lifecycle. We present \toolname, an open-source platform that integrates these activities into a single workflow. 
\toolname represents agents, prompts, tools, models, parameters, and communication topology in exportable, versioned JSON specifications, translated into executable LangGraph workflows supporting different interaction structures. Users can modify and re-evaluate configurations using local or cloud-based LLMs. 
We instantiate \toolname for automated program repair (APR) and evaluate a sequential baseline and seven reconstructed configurations on synthetic and real-world SWE-bench Lite issues. The results reveal substantial cost differences among configurations with similar repair outcomes, while the baseline resolves 12 of 77 SWE-bench issues using open-source LLMs. Additional proof-of-concept tasks illustrate applicability beyond APR. \toolname supports iterative experimentation with heterogeneous MAS configurations, enabling researchers to examine their outcomes and costs within a common environment.
Demonstration video: \url{https://youtu.be/278sOR4lBEk}.
\end{abstract}

\begin{IEEEkeywords}
    Multi-Agent Systems (MAS), Automated Program Repair (APR), Large Language Models (LLMs), SWE-bench 
\end{IEEEkeywords}

\section{Introduction}
Large Language Models (LLMs) are increasingly used in agentic solutions for Software Engineering (SE) tasks, including automated program repair (APR) \cite{hou2024largelanguagemodelssoftware,jimenez2024swebench}. Multi-agent systems (MAS) address these tasks by coordinating specialized agents, such as bug localizers, patch generators, and validators \cite{li2024agentsneed,he2025llm}. How these agents communicate is an important design choice: different topologies determine how information flows between agents and may affect both task outcomes and computational cost \cite{zhuge2024gptswarm,zhang2025gdesignerarchitectingmultiagentcommunication}.

Developing a MAS is only part of the challenge. Researchers investigating alternative architectures also need to adjust configurations, run them on comparable tasks, evaluate their outputs, and track the results. Existing frameworks and visual development environments simplify the construction and execution of agent workflows. However, comparing MAS configurations for SE tasks requires managing benchmark instances and evaluation results. Without an integrated workflow, such experiments become harder to organize and reproduce\cite{de2026prompting}.

We present \toolname, an open-source platform designed to support this experimentation process. \toolname represents agents as nodes and their communication topology as edges, while keeping prompts, tools, models, and parameters configurable. Each system is captured in an exportable, versioned JSON specification that is translated into an executable LangGraph workflow\footnote{\url{https://www.langchain.com/langgraph}}.Researchers can reuse agent definitions, modify their communication structure, and execute alternative configurations using local or cloud-based agents. By storing specifications and results, \toolname enables \textit{researchers} and \textit{practitioners} to reproduce experiments, compare configurations, and iteratively refine MASs.



We demonstrate this workflow through APR, integrating \toolname with the SWE-bench evaluation harness \cite{jimenez2024swebench}. We evaluate a sequential baseline and seven reconstructed configurations on synthetic and real-world bugs using the same locally hosted model. On a 23-instance development subset, the baseline and \code{hyperagent\_star} each resolve two issues, while the baseline requires 39.78\% less wall-clock time and 47.32\% fewer tokens. The baseline also resolves 12 of 77 selected SWE-bench Lite test instances. These experiments illustrate how \toolname can be used to execute and compare heterogeneous MAS configurations under a common evaluation setup. Additional proof-of-concept workflows illustrate its applicability beyond APR.


\resquest{Tool and demonstration video availability}{
The source code and documentation of \toolname, together with the replication artifacts, are available on
GitHub~\faGithub{}~\cite{OPEN_DRACO_REPOSITORY_URL} and
Figshare~\aiFigshare{}~\cite{yepan2026opendraco}. A demonstration video of
\toolname is available on YouTube~\faYoutube{}~\cite{demo-video}.}

\section{Related Work}
The organization of agents and their communication channels is an important design dimension of LLM-based MAS. Early approaches relied on predefined interaction patterns, including sequential pipelines, hierarchical structures, and ensembles. Li et al.~\cite{li2024agentsneed}, for example, investigate how aggregating multiple agents can improve task performance. More recent studies examine communication topology directly. Yang et al.~\cite{yang2025topologicalstructurelearningresearch} identify topology as an important research direction, while GPTSwarm and G-Designer investigate the optimization of agent interaction graphs \cite{zhuge2024gptswarm,zhang2025gdesignerarchitectingmultiagentcommunication}. AFlow instead focuses on automatically generating agent workflows \cite{zhang2025aflow}. These approaches contribute methods for constructing or optimizing MAS designs. Our focus is complementary: providing an environment in which researchers can configure, execute, and evaluate alternative designs for SE tasks.

Several development platforms already provide mechanisms for building and inspecting agent workflows. Flowise~\cite{FlowiseAI}, Langflow~\cite{Langflow}, and Crew Studio~\cite{CrewAI} offer visual interfaces for composing agents and their interactions. AutoGen Studio~\cite{AutoGen} also supports declarative JSON specifications, reusable components, and interactive evaluation, while LangGraph Studio~\cite{langgraph} provides graph visualization and execution debugging. Thus, visual modeling, reusable agents, and declarative workflow specifications are not unique to \toolname.

\toolname differs primarily in its focus on comparative experimentation for SE. Rather than proposing a new topology optimization method or general-purpose workflow builder, it combines configurable agents and communication structures with versioned specifications, execution, task-specific evaluation, and result comparison. Its APR instantiation integrates reusable agents and topology templates with SWE-bench evaluation, allowing researchers to modify, reuse, and compare MAS configurations and their outcomes and costs without reimplementing orchestration logic.

\section{\toolnameplain Architecture}
\begin{figure}[ht]
\centering
\includegraphics[width=\columnwidth]{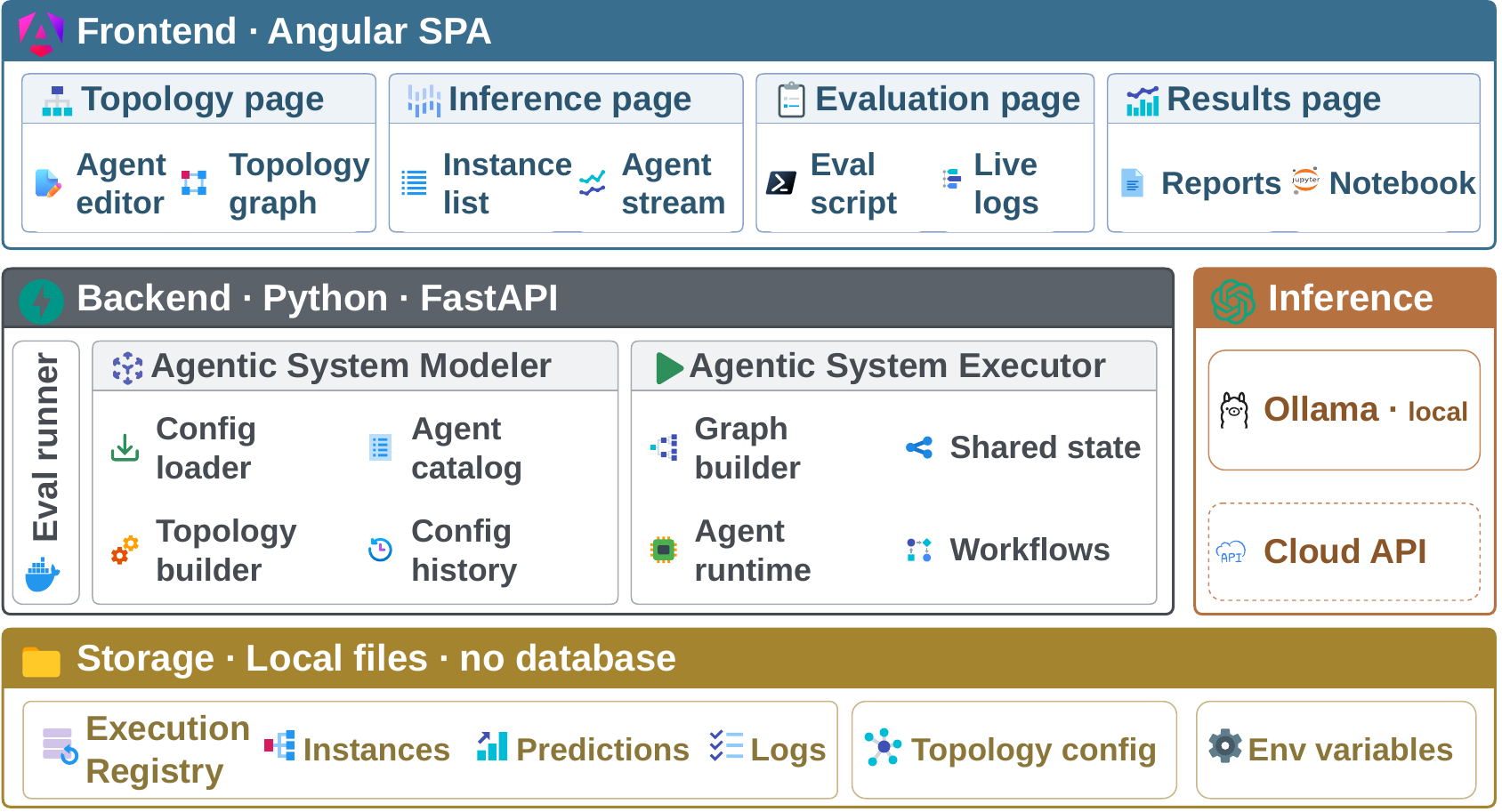}
\caption{Component architecture of \toolname.}
\label{fig:architecture}
\end{figure}

\figref{fig:architecture} shows the architecture of \toolname. The framework
follows an n-tier design. Each tier is described below. 

\subsection{User Workflow}
\label{sec:workflow}

\begin{figure}[ht]
\centering
\includegraphics[width=\columnwidth]{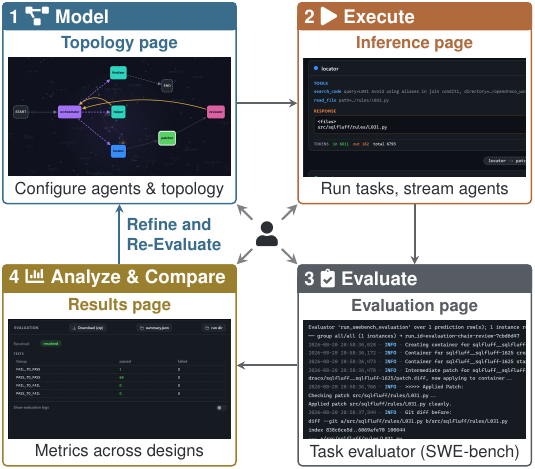}
\caption{\toolname user workflow.}
\label{fig:user-workflow}
\end{figure}

\figref{fig:user-workflow} summarizes the workflow followed by a user when configuring and executing a MAS with \toolname. 
\toolname can be accessed through its web interface or command line and deployed locally through Docker. Users select or define a topology, configure each agent's role, prompt, tools, model, and parameters, and save the configuration for reuse and versioning. They then execute it on selected task instances, evaluate the outputs, and inspect the results. Finally, they can modify or replace the configuration and rerun it on the same instances to compare MAS designs under the same setting.

\subsection{Frontend Layer}
\label{sec:frontend}
\toolname provides an Angular-based interface. The \textbf{Topology} page connects to the Agentic System Modeler (\secref{sec:arch:modeler}), enabling users to define MASs as agent nodes and communication edges and configure their models, prompts, parameters, and tools.
Configurations are stored as importable/exportable JSON files and Git-versioned. The \textbf{Inference} page executes selected instances, streams agent activity, and stores outputs. The \textbf{Evaluation} page runs task-specific evaluators; for APR, it uses the \swebench harness to apply patches and execute tests. The \textbf{Results} page reports outputs, patches, evaluation results, wall-clock time, token usage, passed tests, and resolution status.

\subsection{Backend Layer}
The backend is implemented in Python and exposes its services through a FastAPI
server, which is consumed by the frontend and can also be accessed by third-party components.

\subsubsection{Agentic System Modeler}
\label{sec:arch:modeler}
This module supports the modeling of MAS and produces a structured specification (e.g., JSON) containing:
\begin{inparaenum}[\it a)]
    \item each agent, defined by its model, prompts, parameters (e.g., temperature),
    and tools; and
    \item the topology, i.e., how the agents are connected.
\end{inparaenum}
It further provides:
\begin{inparaenum}[\it i)]
    \item {\bf a validation mechanism}, which checks the specification against LangGraph's constraints and detects structural anomalies (e.g., unreachable agents or the absence of a path from \code{START} to \code{END}). Cyclic flows are supported and bounded at runtime by a revisit limit; 
    \item {\bf an agent catalog}, which organizes agents by attributes such as role to support reuse across different systems;
    \item {\bf configuration history}, which maintains a Git history for each configuration file, allowing changes to be tracked and previous versions to be restored; and
    \item {\bf persistence}, which stores specifications in the file system and loads them when needed for reuse or sharing.
\end{inparaenum}

\subsubsection{Agentic System Executor}
This module converts a modeled MAS, represented as a JSON specification (\secref{sec:arch:modeler}), into a LangGraph \code{StateGraph} and runs it on a given problem instance to produce a prediction (e.g., a patch for a \swebench instance).
It comprises:
\begin{inparaenum}[\it i)]
    \item a graph builder that creates the LangGraph representation;
    \item a typed shared state for agent communication;
    \item an agent runtime that calls the model with its configured parameters,
    prompt, tools, and current state; and
    \item a workflow runner that schedules nodes and supports fan-out and
    conditional edges.
\end{inparaenum}
Each agent follows a common {\bf reasoning-and-acting} loop~\cite{yao2023reactsynergizingreasoningacting}, alternating between LLM calls and tool invocations until it produces a final response. Outputs are normalized before being passed downstream. A shared registry exposes the tool catalog and enforces per-agent allow lists defined by the \code{tools} list.
\subsubsection{Evaluation Runner}
\label{sec:arch:evaluator}
This module evaluates outputs produced by the \emph{Agentic System Executor}, including correctness or quality against ground-truth solutions. 
For APR, \toolname provides three modes: 
\begin{inparaenum}[\it i)]
   \item the official \swebench harness in Docker, 
   \item remote execution via \code{sb-cli}, and 
   \item a local evaluator for repositories outside \swebench{}---including those containing our synthetic instances---that clones the repository, applies the generated patch, and runs its tests.
\end{inparaenum}

\subsubsection{Additional Layers}
The \emph{Storage Layer} persists execution registries, predictions, reports, metrics, configurations, logs, and results as local files, enabling direct inspection and editing. 
The \emph{Inference Layer} supports Ollama\footnote{\url{https://ollama.com}} backends running on \code{localhost} (the default) or another host, and cloud-based services through OpenAI-compatible APIs (e.g., OpenRouter).

\subsection{Customization of \toolname for APR}

We instantiate \toolname for APR by specializing three components:
\begin{inparaenum}[\it i)]
   \item The {\bf agent catalog}, comprising agents derived from 16 publicly available repositories identified in our previous study~\cite{martinez2026s}. These repositories implement solutions reported on the \swebench leaderboard, and the agents are classified by role (e.g., \emph{Bug Localizer} or \emph{Patch Synthesizer}).
    The catalog includes only artifacts covered by permissive upstream licenses (MIT, BSD, or Apache-2.0) and records their provenance.
    \item The {\bf evaluation pipeline}, through integration with the \swebench harness. 
    \item The {\bf UI}, which exposes the APR agent catalog and predefined APR
    topology templates, provides access to the \swebench evaluation, and
    reports APR-specific results and metrics.
\end{inparaenum}

\section{Evaluation and  Preliminary Results}
\subsection{Evaluation Goals} Our preliminary evaluation addresses four questions: whether \toolname can represent and execute existing agentic APR topologies end-to-end; whether the modeled systems can repair real-world bugs from \swebench Lite; whether the evaluation pipeline scales to a larger set of real-world instances; and how temperature affects run-to-run variability. We use synthetic bugs for the initial end-to-end assessment.

\subsection{Evaluation Metrics and Setup}
\label{sec:eval:setup}
We use three instance sets: five synthetic APR instances spanning five difficulty tiers, the 23-instance \code{dev} subset of \swebench Lite, and a 77-instance subset of the \swebench Lite test split, restricted to instances whose gold patch is shorter than 600 characters. The synthetic set allows us to evaluate all eight configurations end-to-end before running the more expensive real-world experiments.

We evaluate eight MAS configurations that vary in their topology and/or agent definitions. Collectively, they cover four topology patterns: \code{chain} (sequential), \code{star} (via a central agent), \code{tree} (hierarchical), and \code{hybrid} (combined). 
First, we define a \code{chain} baseline following a conventional APR sequence by connecting a bug-localization agent, a patch-generation agent, and a validation agent.

The remaining seven configurations draw on agentic APR systems reported on the \swebench leaderboard. Two configurations, \code{tree} and \code{hybrid}, are inspired by the communication structures of Prometheus and AgentScope, respectively, while using our own agent definitions. 
The other five reconstruct systems by combining agents from the catalog according to their communication structures.
To assess run-to-run variability, we execute the \code{chain} baseline five times at each temperature (0/1) on the synthetic set, using a different random seed for each run and recording the resolved rate, wall-clock time, and token usage.

\begin{table}[ht]
    \centering
    \caption{Results with \code{qwen3.5:9b}. \textbf{Ag.\,(ed.)}: agents (edges); \textbf{Resolved}: synthetic difficulty from trivial to expert.
    }
    \label{tab:swelite}
    \footnotesize
    \setlength{\tabcolsep}{4pt}
    \resizebox{\columnwidth}{!}{%
        \begin{tabular}{@{}lcrcl@{}}
            \toprule
            \textbf{Configuration}                                                                     & \textbf{Ag.\,(ed.)}
                                                                                                       & \textbf{Wall-clock}                           & \textbf{Resolved}
                                                                                                       & \textbf{Total tokens}                                                                                             \\
            \midrule
            \multicolumn{3}{@{}l}{\emph{Configuration comparison: synthetic set (5 difficulty tiers)}} & {\scriptsize\makebox[7.5ex][c]{$-$\hfill$+$}} &                                                                   \\
            \code{chain} (baseline)                                                                    & 4\,(3)                                        & 29\,min\,2\,s         & \resolvedboxes{11011}
                                                                                                       & 226{,}605                                                                                                         \\
                                                                                                       
            \code{tree}                                                                    & 7\,(6)                                        & 31\,min\,41\,s    & \resolvedboxes{11010}
                                                                                                       & 172{,}084                                                                                                         \\
            \code{hybrid}                                                                  & 6\,(7)                                        & 35\,min\,21\,s    & \resolvedboxes{11011}
                                                                                                       & 254{,}316                                                                                                         \\
            \code{experepair\_star}                                                                    & 5\,(7)                                        & 41\,min\,18\,s    & \resolvedboxes{10011}
                                                                                                       & 197{,}781                                                                                                         \\
            \code{hyperagent\_star}                                                                    & 5\,(7)                                        & 29\,min\,35\,s    & \resolvedboxes{11111}
                                                                                                       & 229{,}188                                                                                                         \\
            \code{joycode\_star}                                                                       & 5\,(7)                                        & 39\,min\,15\,s    & \resolvedboxes{11111}
                                                                                                       & 256{,}908                                                                                                         \\
            \code{lingxi\_star}                                                                        & 5\,(7)                                        & 42\,min\,22\,s    & \resolvedboxes{11111}
                                                                                                       & 300{,}563                                                                                                         \\
            \code{openhands\_star}                                                                     & 6\,(9)                                        & 17\,min\,32\,s    & \resolvedboxes{11111}
                                                                                                       & 82{,}840                                                                                                          \\
          
            \midrule
            \multicolumn{5}{@{}l}{\emph{Real-world repair: \swebench Lite (23-instance dev subset)}}                                                                                                                       \\
            \code{chain} (baseline)                                                                    & 4\,(3)                                        & 4\,h\,37\,min     & 2/23
                                                                                                       & 2{,}178{,}309                                                                                                     \\
            \code{hyperagent\_star}                                                                    & 5\,(7)                                        & 7\,h\,40\,min     & 2/23
                                                                                                       & 4{,}134{,}877                                                                                                     \\
            \code{openhands\_star}                                                                     & 6\,(9)                                        & 3\,h\,15\,min     & 0/23
                                                                                                       & 1{,}348{,}773                                                                                                     \\
            \midrule
            \multicolumn{5}{@{}l}{\emph{Evaluation scalability: \swebench Lite (77-instance test subset)}}                                                                                                                 \\
            \code{chain}                                                                               & 4\,(3)                                        & 16\,h\,4\,min     & 12/77
                                                                                                       & 6{,}344{,}990                                                                                                     \\
            \midrule
            \multicolumn{5}{@{}l}{\emph{Run-to-run variability: \code{chain} on the Synthetic set}}                                                                                                                        \\

            \code{chain} - temperature 0                                                               & 4\,(3)                                        & 30--32\,min       & 3/5 (all)
                                                                                                       & 206.0--206.5k                                                                                                   \\
            \code{chain} - temperature 1                                                               & 4\,(3)                                        & 25--34\,min       & 2/5 to 4/5
                                                                                                       & 198.7--276.8k                                                                                                   \\
            \bottomrule
        \end{tabular}%
    }
\end{table}

\subsection{Preliminary Results}

\tabref{tab:swelite} reports the results. All configurations use the same model
(\code{qwen3.5:9b}) to ensure a consistent comparison. On the synthetic set, \code{openhands\_star} resolves all five instances and also has the shortest wall-clock time and lowest token usage among the eight configurations. On the 23-instance development subset, both \code{chain} and \code{hyperagent\_star} resolve two instances. However, \code{chain} uses 39.78\% less time and 47.32\% fewer tokens. Given this effectiveness--cost trade-off, we select \code{chain} for the
larger 77-instance test subset. It resolves 12 of the 77 instances (15.58\%). 

At temperature~0, \code{chain} resolves the same three synthetic instances in
all five runs. At temperature~1, the number of resolved instances ranges from
two to four, and both runtime and token usage vary more across runs. This variability suggests that configurations using nonzero temperature should
be evaluated over repeated runs.

%
%

\subsection{Discussion}

\subsubsection{Framework Extensibility}
To assess \toolname beyond APR, we modeled and executed proof-of-concept workflows through its UI (Section~\ref{sec:frontend}), including a translation
workflow that reads a text file, translates it, and evaluates the output with BLEU against a reference. This required only task-specific agents, topologies, tools, and evaluators, without changes to the framework core. \toolname can support new tasks through configuration without framework
modification.

\subsubsection{Limitations of the Experiments}

\begin{description}[leftmargin=0.3cm]

\item \underline{\textsl{Reconstruction Fidelity:}}
The reconstructed APR configurations approximate the original systems in terms
of agents, prompts, tools, and communication structures, but not their exact
runtime behavior. Thus, our results characterize their \toolname instantiations,
and differences cannot be attributed to topology alone.

\item \underline{\textsl{Resources:}}
Due to the 6 GB VRAM of the GPU, all agents were executed sequentially, including
independent branches. Reported wall-clock times therefore do not capture
potential parallelization gains.

\item \underline{\textsl{Model Choice:}}
All experiments use the locally hosted \code{qwen3.5:9b}, limiting
generalizability to other models. We chose an open local model to support
reproducibility, keep source code local, and avoid proprietary paid APIs.

\item \underline{\textsl{Evaluation Scope:}}
Except for the variability study, each configuration was executed once.
Moreover, only three configurations were evaluated on real-world instances,
with \code{chain} alone evaluated on the larger subset, limiting conclusions
about variability and configuration differences.

\end{description}
\section{Conclusion and Future Work}
This paper presented \toolname, an open-source platform for modeling, executing,
and evaluating configurable LLM-based MAS for SE. \toolname captures agents, prompts, tools, models, parameters, and communication topology in a versioned, shareable configuration, allowing MAS experiments to be modified and reproduced without changing the framework core. A preliminary APR evaluation shows that \toolname supports end-to-end experimentation and comparison of configurations by effectiveness, time, and token usage. Future work will support runtime topology evolution, evaluate additional local and cloud-based models over repeated runs, and extend the agent catalog, topology templates, and evaluators to other SE tasks. We also plan a user study on modeling and reproducibility effort, and to incorporate sustainability metrics into \toolname.

\section*{Acknowledgements}
ChatGPT was used only to improve language and readability. All authors reviewed and validated the final manuscript and take full responsibility.

\bibliographystyle{IEEEtran}
\bibliography{bib}

\end{document}